\documentstyle [twocolumn,epsf]{mn}
\newcommand{\mincir}{\raise
-3.truept\hbox{\rlap{\hbox{$\sim$}}\raise4.truept\hbox{$<$}\ }}
\newcommand{\magcir}{\raise
-3.truept\hbox{\rlap{\hbox{$\sim$}}\raise4.truept\hbox{$>$}\ }}
\newcommand{\minmag}{\raise
-3.truept\hbox{\rlap{\hbox{$<$}}\raise5.truept\hbox{$<$}\ }}
\newcommand{\be}{\begin{equation}}
\newcommand{\ee}{\end{equation}}
\newcommand{\ba}{\begin{eqnarray}}
\newcommand{\ea}{\end{eqnarray}}
\newcommand{\brr}{\begin{array}}
\newcommand{\err}{\end{array}}
\newcommand{\bc}{\begin{center}}
\newcommand{\ec}{\end{center}}

\title[PSCz Voids]{The Size and Shape of Local Voids}
\author[Plionis \& Basilakos]
{Manolis Plionis$^{1}$ \& Spyros Basilakos$^{2}$ \\
\vspace{0.1cm}
$^1$ Institute of Astronomy \& Astrophysics, National Observatory of Athens, 
I.Metaxa \& B.Pavlou, Palaia Penteli, 152 36, Athens, Greece \\
$^2$ Astrophysics Group, Imperial College London, Blackett Laboratory, 
Prince Consort Road, London SW7 2BW, UK\\
}

\begin{document}

\maketitle

\begin{abstract}
We study the size and shape of low density regions 
in the local universe which we identify in the smoothed 
density field of the PSCz
flux limited IRAS galaxy catalogue. After quantifying the systematic biases 
that enter in the detection of voids using our data set and method,
we identify, using a smoothing length of 5 $h^{-1}$ Mpc, 14 voids
within 80 $h^{-1}$ Mpc (having volumes $\ge 10^{3}$ $h^{-3}$ Mpc$^{3}$)
and using a smoothing length of 10 $h^{-1}$ Mpc, 8 voids within 130
$h^{-1}$ Mpc (having volumes $\ge 8 \times 10^{3}$ $h^{-3}$ Mpc$^{3}$).
We study the void size distribution and morphologies and 
find that there is roughly an equal number
of prolate and oblate-like spheroidal voids. 
We compare the measured PSCz void shape and size distributions
with those expected in six different CDM models and find that
only the size distribution can discriminate between models. The
models preferred by the PSCz data are those with intermediate values
of $\sigma_{8} (\simeq 0.83)$, independent of cosmology.

\end{abstract}

\section{Introduction}
Many authors have claimed that low density regions (voids) are the most common 
features of the large scale structure 
of the universe owing to the fact that they occupy more than a half of 
its volume. Individual voids and their properties 
have been investigated by different authors (cf. Jo\~eveer et al. 1978;
Kirshner et al.1981; Rood 1981 and references that he gives). 
Since their creation has been attributed
to the effects of gravitational instability 
(cf. Zeldovich, Einasto \& Shandarin 1982; 
Coles, Melott \& Shandarin 1993; Peebles 2001 and 
references therein), the distribution of voids
on large scales could provide important constraints on 
models of structure formation. 
From N-body simulations it has become evident that in 
the hierarchical structure formation scenario, 
matter collapses and forms high density objects, 
like clusters and superclusters,
and enhance the underdense regions between them (cf. Melott et al. 1983). 
Therefore, if the above view is correct then the
morphological and statistical properties of voids (size, shape) 
should depend on the initial power 
spectrum $P(k)$ and on the density parameter $\Omega_{\circ}$ 
(cf. White 1979; Melott 1987; Einasto, Einasto \& Gramann 1989;
Reg\"os \& Geller 1991; Ryden \& Melott 1996). 
Blaes, Goldreich \& Villumsen (1990) and Van de Weygaert \& Van Kampen (1993) 
studied the morphological properties of voids and found, that the latter 
structures become nearly spherical in their very underdense internal regions. 

In order to study in an objective manner the distribution of voids and
their physical properties it is necessary to develop objective
void-finding algorithms and to apply them onto well controlled data.
Such attempts were pioneered by Kauffmann \& Fairall (1991),
Kauffmann \& Melott (1992) and more recently by
El-Ad, Piran \& da Costa (1996), El-Ad \& Piran (1997) and Stavrev (2000).  
El-Ad, Piran \& da Costa (1997) applied their void-finding algorithm
to the IRAS 1.2Jy galaxy catalogue and found, within 80 $h^{-1}$ Mpc,
15 voids with an average diameter of $40\pm 6 h^{-1}$Mpc.
M\"uller et al. (2001) have also studied the distribution 
of voids and their sizes
using the large Las Campanas Redshift Survey and compared their
properties with CDM simulations. They found that although the void-size
distribution provides important information on the large-scale
distribution of matter, galaxy biasing seems more important in defining
voids than differences between the cosmological models. 

In this paper we use the PSCz-IRAS redshift survey 
in order to measure the void size and shape distributions in the
local universe
and to investigate whether these distributions can be used as 
cosmological probes.
The plan of the paper is the following:
In section 2 we briefly describe the PSCz data. In section
3 we present the void identification and shape determination procedure,
the systematic effects that affect the PSCz void detection and our
results. We compare our PSCz results with the corresponding ones of six
cosmological models in section 4 and finally in section 5 
we draw our conclusions.

\section{The PSCz galaxy sample}
In our analysis we use the recently completed IRAS flux-limited 
60-$\mu$m redshift survey (PSCz) which is described in
Saunders et al. (2000). The PSCz 
catalogue contains $\sim 15500$ galaxies with flux $S_{lim}\ge 0.6$ Jy
covering the $\sim 84\%$ of the sky. 

To construct an unbiased 
continuous density field we have to 
take into account the well-known degradation of
sampling as a function of distance due to the fact that the PSCz catalogue is 
a flux limited sample. This is done
by weighting each galaxy by the inverse selection function,
assuming that the fainter unobserved galaxies are spatially
correlated with the bright observed ones.
Note that the selection
function is defined as the fraction of the galaxy number density 
that is observed above the flux limit at some distance $r$. 
We estimate the selection function by using the functional 
form of the IRAS galaxy luminosity function 
of Saunders et al. (1990) with parameters: 
$L_{*}=10^{8.45} \; h^{2} L_{\odot}$, $\sigma=0.711$, $\alpha=1.09$ 
and $C=0.0308$ (Rowan-Robinson et al. 2000).  

In order to treat the 16\% excluded sky 
(galactic plane, high cirrus emission areas and unobserved regions) 
we use the PSCz data reduction of Branchini et al. (1999), in which they
followed the Yahil et al. (1991) method to fill the galactic plane
region. This method consists of dividing the Galactic strip into 36
bins of 10$^{\circ}$ in longitude and in distance bins of 1000 km/sec
width. The galaxy distribution in each bin is then simulated by random
sampling the adjacent galactic latitude and distance strips. 
Therefore the local
density fluctuations are extrapolated in the galactic plane. In the
high Galactic latitudes, the unobserved regions are filled
homogeneously with simulated galaxies following the PSCz 
radial selection function.

Finally, in this kind of analysis it is essential to transform redshifts 
to 3D distances in order to minimise the so called ``Kaiser'' effect. 
This effect can be understood by noting that the distribution
of galaxies in redshift space is a distorted representation of
that in real comoving space due to their peculiar velocities 
(Kaiser 1987). In this paper we use 
the 3D distances determined by the iterative algorithm of 
Branchini et al (1999) in which they assume that (a) peculiar velocities
are produced by gravitational instability (b) non-linear effects can be
neglected once the density field is smoothed and (c) galaxies and mass
fluctuations are related through a linear biasing relation
with $\beta=0.5$\footnote{$\beta=\Omega_{\circ}^{0.6}/b_{\rm IRAS}$, with 
$b_{\rm IRAS}$ the IRAS bias factor. 
Our results remain qualitatively the same also for $\beta=1$.},
This value of $\beta$ is strongly suggested from the Branchini et al. (2001)
comparison between the PSCz density field and 
the SFI galaxy velocities (cf. Haynes et al. 1999).

\section{PSCz Voids}
\subsection{Smoothing Procedure}
For the purpose of this study we need to derive from the discrete
distribution of PSCz galaxies a smooth continuous density field. This
is realized by utilizing a Gaussian kernel on a $N^{3}$ grid and
using two smoothing radii, namely 
$R_{{\rm sm}} = 5 \; h^{-1}$ Mpc and 10 $h^{-1}$ Mpc, to probe 
essentially different void sizes. 

As we verified in Basilakos, Plionis \& Rowan-Robisnon (2001),
the coupling between the selection function and the constant 
radius smoothing, results in a distorted smoothed
density distribution, especially at large distances.
Gaussian spheres, centered on distant cells, 
overestimate the true density in regions where galaxies are
detected (due to the heavy selection function weighting), 
while in underdense regions 
they underestimate the true density. Using numerical simulations,
Basilakos et al. (2001) identified the depth out to which these
distortions are relatively small and 
developed a phenomenological approach to correct such biases. 
Their procedure
is effective in recovering especially the high-density end of the 
probability density function ({\em pdf}).

In the present study we use such a corrected density field 
and based on the above analysis we extend our
smoothing procedure out to $r_{max}$, where $r_{max}=130$ and 170 $h^{-1}$ Mpc
respectively for the two smoothing radii while the size of each
cell is set equal to $R_{{\rm sm}}$. We will investigate further
possible systematic effects of our method and data by using a Monte-Carlo 
approach in section 3.4.

\subsection{Void Detection}
In order to find our void candidates we select 
all grid-cells with $|b|\ge 10^{\circ}$ and with 
an overdensity under a chosen threshold and 
then join together those having common boundaries.
Due to the fact that voids should be identified 
as low density regions,
the threshold value of the overdensity ($\delta_{\rm th}$) could be 
defined directly from the probability density function 
as its 10$\%$-ile value ($\delta_{th}=-0.68$ and 
-0.39 for the $R_{sm}=5 h^{-1}$Mpc 
and $R_{sm}=10 h^{-1}$Mpc respectively). Using higher 
values of $\delta_{\rm th}$ we tend to connect voids and
percolate underdense regions through the 
whole volume. Finally, we choose to analyse the above candidate voids 
for two different limiting distances, 
namely $r_{\rm lim}\simeq 80 h^{-1}$Mpc for 
$R_{\rm sm}=5$ $h^{-1}$ Mpc, and 
$r_{\rm lim}\simeq 130$ $h^{-1}$ Mpc for 
$R_{\rm sm}=10$ $h^{-1}$ Mpc
respectively in order to avoid distance dependent systematic effects
(see section 3.4 and figure 1).

\subsection{Shape Statistics}
Shapes are estimated for those ``voids'' that 
consist of 8 or more connected underdense cells, utilizing the moments of
inertia ($I_{ij}$) method to fit the best triaxial
ellipsoid to the data (cf. Carter \& Metcalfe 1980; 
Plionis, Barrow \& Frenk
1991; Basilakos et al. 2001). We diagonalize the inertia tensor 
\begin{equation}\label{eq:diag}
{\rm det}(I_{ij}-\lambda^{2}M_{3})=0 \;\;\;\;\; {\rm (M_{3} \;is \; 
3 \times 3 \; unit \; matrix) } \;,
\end{equation}
obtaining the eigenvalues $\alpha_{1}$, $\alpha_{2}$, 
$\alpha_{3}$ (where $\alpha_{1}$ is the semi-major axes) 
from which we define the shape of the configuration since, the
eigenvalues are directly related to the three principal axes 
of the fitted ellipsoid. The volume of each void is then 
$V=\frac{4\pi}{3} \alpha_{1} \alpha_{2} \alpha_{3}$.   
According to the logic of our void-finder, 8 underdense cells
in a row can be connected to form a 'void', which would then appear as
purely prolate. 

The shape statistic procedure, that we use, is based on a differential
geometry approach, introduced by Sahni et al (1998) 
[for application to astronomical data see 
Basilakos et al. 2001]. Here we review only some basic concepts.
A set of three shapefinders are defined having dimensions of length;
${\cal H}_{1}=V S^{-1}$, ${\cal H}_{2}=S C^{-1}$ and ${\cal H}_{3}=C$, 
with $S$ the surface area and $C$ the integrated mean curvature.
Then it is possible to define a set of two dimensional shapefinders $K_{1}$ and
$K_{2}$, as: 
\be
K_{1}=\frac{ {\cal H}_{2}-{\cal H}_{1} }{ {\cal H}_{2}+{\cal H}_{1} } 
\ee
and
\be
K_{2}=\frac{ {\cal H}_{3}-{\cal H}_{2} } { {\cal H}_{3}+{\cal H}_{2} } \;\;, 
\ee
normalized to give ${\cal H}_{i}=R$ ($K_{1,2}=0$) for a sphere of radius $R$.
Therefore, based on these shapefinders we can characterise the morphology of
cosmic structures (underdense or overdense regions) 
according to the following categories: (i) oblate-like ellipsoids for $K_{1}/K_{2}>1$;
(ii) prolate-like ellipsoids for $K_{1}/K_{2}<1$; (iii) triaxial for $K_{1}/K_{2}\simeq 1$ and (iv)
spheres for $\alpha_{1} \simeq \alpha_{2} \simeq \alpha_{3}$ and thus $(K_{1},K_{2}) \simeq (0,0)$.
For the quasi-spherical objects the ratio $K_{1}/K_{2}$ 
measures the deviation from pure sphericity.

\subsection{Test for systematic errors}

We run a large number of Monte-Carlo simulations in 
which we destroy the intrinsic PSCz galaxy clustering 
by randomising the angular coordinates of the galaxies while 
keeping their distances and therefore their selection function unchanged. 
On this intrinsically random galaxy distribution, 
utilizing the procedure described before, we identify the expected
random voids, $N_{\rm rand}$, which are due to our void-identification
method itself.
In figure 1 we plot for the different smoothing radii, 
the probability of detecting real voids in the PSCz data, defined as
${\cal P} = 1-N_{\rm rand}/N_{\rm PSCz}$, 
as well as the number of real PSCz voids, $N_{\rm
PSCz}$. In the presence of negligible systematic biases in our
method and data, the above selection process should result in ${\cal P}
\simeq 1$, at all distances. 

\begin{figure}
\mbox{\epsfxsize=8.8cm \epsffile{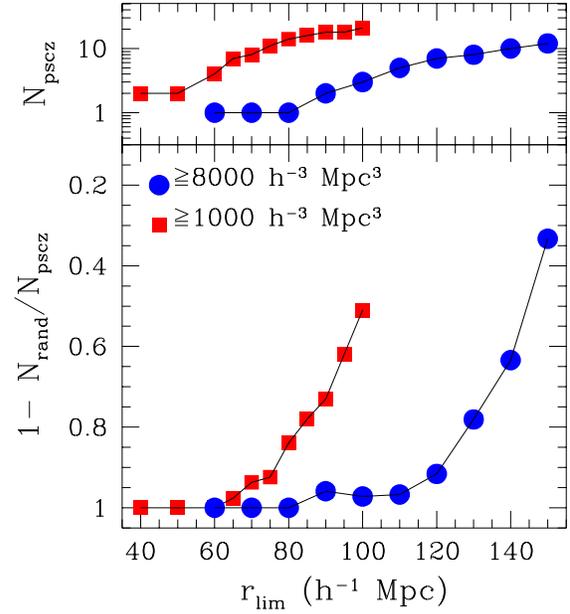}}
\caption{Statistical significance of our void detection procedure as a
function of distance 
using either $R_{\rm sm}=5 h^{-1}$Mpc or $R_{\rm sm}=10h^{-1}$Mpc.}
\end{figure}
We find that the number of random voids increases 
with distance, a fact which should be attributed
to the coupling of the PSCz selection function (ie., the degradation of
the observed mean galaxy density as a function of distance which
produces an inherently noisier density field at large distances) 
and the constant radius smoothing. 
Our phenomenological method of recovering the 
true density field (Basilakos et al. 2001) 
is multiplicative in nature and thus it is unable 
to correct the low and zero density cells. Therefore, the lower 
part of the {\em pdf} cannot be recovered efficiently. 
In particular we find that for $R_{\rm sm}=5h^{-1}$Mpc 
smoothing radius (figure 1, squares), we have ${\cal P}>0.8$
for detecting real voids within $r \le 80^{-1}$Mpc 
while for the $R_{\rm sm}=10 h^{-1}$Mpc smoothing radius (figure 1,
circles), a similar probability is found within $r \le 130^{-1}$Mpc. 
Note that for the largest voids in our sample this probability is much
larger ($>$0.98).

\subsection{The PSCz Void Cosmography}

In figure 2 we plot the smoothed PSCz galaxy distribution on the
supergalactic plane out to 90 $h^{-1}$ Mpc. The contour 
step is $\Delta\delta=0.6$, starting from $\delta_{th} =-0.68$)
while the $\delta=0$ level appears as a thick continuous 
line. All major known clusters are present, like the Virgo cluster at
$X_{sup}\simeq -5$, $Y_{sup}\simeq 5$; the Hydra-Centaurus at
$X_{sup}\simeq 35$, $Y_{sup}\simeq 15$; the Perseus at 
$X_{sup}\simeq 50$, $Y_{sup}\simeq -20$ and the Coma at
$X_{sup}\simeq 0$, $Y_{sup}\simeq 70$. We also plot the voids that we have
detected as big circles while as crosses we mark the IRAS 1.2Jy voids found by El-Ad et
al. (1997). 
We observe that our respective voids have almost identical positions. 
However, we do not detect
two IRAS 1.2Jy voids, which in one case is due to a slight difference in the 
void $Z_{sup}$-position (our void No5 in table 1) and in the other it
is probably due to our adopted overdensity threshold. Furthermore, the
underdense region at the upper right section of the density map extends to
higher $Z_{sup}$ and constitutes a large void (our No 12).

\begin{table*}
\caption[]{Detected voids in the $R_{sm}=5 \;h^{-1}$ Mpc density
field for $\delta \le -0.69$ (corresponding to the 10$\%$-ile) 
with $V\ge 10^{3} \; h^{-3}$ Mpc$^{3}$ ($N_{\rm cell} \ge 8$) and $r
\le 80 \; h^{-1}$ Mpc. The different columns are self-explanatory with
column 3 being the distance of the void center and column 10 the
void semi-major axis. The Local Void is No13 and the Sculptor Void is No4.}
\tabcolsep 7pt
\begin{tabular}{cccccccccc} \hline 
N& Volume ($10^{3} \;h^{-3}$ Mpc$^{3}$) &d ($h^{-1}$ Mpc)&$X_{sup}$ & $Y_{sup}$ & 
$Z_{sup}$ & $K_{1}$ & $K_{2}$ & $K_{1}/K_{2}$& $\alpha_{1}$
($h^{-1}$Mpc)\\ \hline
  1 &       8.6 &  77.4 & 14.8&  -73.2  &-20.6&        0.040 &      0.060&   0.67 &   20.7\\
  2  &      2.4 &  72.6 &-6.9&  -58.4 &  42.5 &        0.073 &      0.205&   0.35 &   19.9\\
  3  &     12.6 &  78.5 & 69.6&  -28.9 &  22.0 &       0.033 &      0.045&   0.72 &   22.1\\
  4  &     62.0 &  78.2 &-67.4&  -39.5 &  -4.8  &      0.039 &      0.042&   0.94 &   36.8\\
  5  &      3.8 &  33.9 &-13.8&  -24.9 & -18.4   &     0.017 &      0.014&   1.21 &   12.0\\
  6   &     1.8 &  73.9 & 27.6 & -28.2  & 62.5     &   0.070  &     0.041 &  1.70  &  10.9\\
  7    &    2.2 &  38.6 &-17.2&  -16.7 &  30.3      &  0.021 &      0.031&   0.67 &   11.6\\
  8      &  8.9 &  70.0 & 13.6&   34.1 & -59.6  &      0.028&       0.024&   1.17&    17.3\\
  9       & 1.4 &  63.4 &-42.9&   33.3 & -32.7   &     0.078&       0.286&   0.27&    19.5\\
 10  &      1.2 &  50.4 & 18.5 &  46.5  & -6.0      &  0.102 &      0.063 &  1.61 &   11.1\\
 11  &      1.8 &  60.6 & 39.0 &  45.1  &-11.1       & 0.033 &      0.032 &  1.03 &   10.6\\
 12  &     16.4 &  64.5 & 37.8 &  49.4  & 17.1&        0.084 &      0.101 &  0.83 &   29.8\\
 13   &    25.4 &  51.2 &-10.6&   30.3&   39.9&        0.079&       0.042&   1.86&    26.6\\
 14    &    1.9 &  71.6 &-8.8  & 63.2&   32.5  &      0.029&
0.034 &  0.85&    11.1\\ \hline
\end{tabular}
\end{table*}

\begin{table*}
\caption[]{Detected voids in the $R_{sm}=10 \;h^{-1}$ Mpc density
field for $\delta \le -0.39$ (corresponding to the 10$\%$-ile) 
with $V\ge 8\times 10^{3} \; h^{-3}$ Mpc$^{3}$ ($N_{\rm cell} \ge 8$) and $r \le 130 \; h^{-1}$ Mpc.}
\tabcolsep 7pt
\begin{tabular}{cccccccccc} \hline
N& Volume ($10^{4} \;h^{-3}$ Mpc$^{3}$) &d ($h^{-1}$Mpc)&$X_{sup}$ & $Y_{sup}$ & 
$Z_{sup}$ & $K_{1}$ & $K_{2}$ & $K_{1}/K_{2}$& $\alpha_{1}$
($h^{-1}$Mpc) \\ \hline
  1 &       2.5  & 129.6&-48.8 &-115.3  & 33.5 &       0.063  &     0.050&   1.27  &  28.1\\
  2   &    63.8  &  117.3&37.8 &-107.4 & -27.9 &       0.077  &     0.048&   1.60 &   81.7\\
  3    &    1.2  & 58.5&-28.5 & -50.9 &   5.0 &       0.067  &     0.059&   1.14 &   23.0\\
  4     &   0.9  &  103.6&72.8 & -38.4  & 62.9 &       0.063  &     0.153&   0.41 &   27.6\\
  5      &  1.6  &  109.4&86.8 & -23.3 & -62.2 &       0.016  &     0.016&   1.00 &   19.9\\
  6       & 7.7 &  97.8&-80.3 & -43.2&   35.4 &       0.063  &     0.193&   0.33 &   61.7\\
  7     &2.8    &111.3&-6.4   &72.9  &-83.9    &    0.048     &  0.044  & 1.09    &28.5\\
  8 &      34.0&  86.8&53.1 &  46.7 &  50.2 &       0.055 &      0.155&
0.36&    92.6\\ \hline
\end{tabular}
\end{table*}

Details for those voids that consist of 8 or more cells are 
presented in table 1 and 2 for both smoothing radii;
the different columns being self explanatory. Evidently 8 voids out of 14 are prolate-like 
ellipsoids and the rest oblate-like. While from table 2, we see that 
we have a slight excess of oblate-like voids, with one purely triaxial void
($K_{1}/K_{2}\simeq 1$). 
Regarding extreme shaped voids, we have found, in the $R_{sm}=5$
$h^{-1}$Mpc field, two very prolate voids (see table 1: void
2 and 9), while in the $R_{sm}=10$ $h^{-1}$ Mpc field,
we again find two such prolate-like voids 
(see table 2: voids 6 and 8). 

The median value of the semi-major void axes is $\sim20$ $h^{-1}$ Mpc and 
$\sim28$ $h^{-1}$ Mpc for $R_{sm}=5$ $h^{-1}$ Mpc and $R_{sm}=10$
$h^{-1}$ Mpc fields respectively, in good agreement with the results of El-Ad
et al (1997). 

If we had used the redshift space PSCz density field we would have
found the same voids, with a similar size distribution 
but biased towards prolate shapes.

\begin{figure}
\mbox{\epsfxsize=8.cm \epsffile{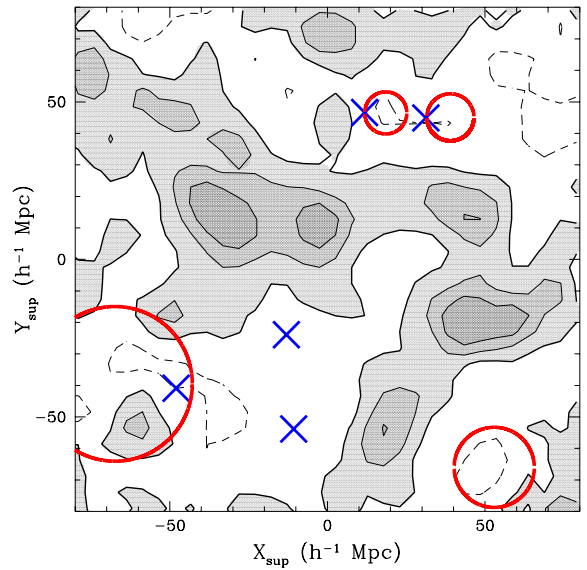}}
\caption{The smooth PSCz density field on the Supergalactic Plane out
to 90 $h^{-1}$Mpc.
Large circles represent our detected voids (with sizes 
proportional to their volume) and crosses represent those of El-Ad et
al. (1997) (with $-17<Z_{sup}<17$ $h^{-1}$).}
\end{figure}

\section{Comparison with Cosmological Models}
We use mock PSCz catalogues (Branchini et al. 1999)
generated from six large cosmological N-body
simulations of Cole et al. (1998), in order to investigate whether void properties
(shapes and numbers) can discriminate between models. We consider six 
different cold dark matter models, which are presented in table 3.

\begin{table}
\caption[]{Model parameters. The
first four are normalized by the observed cluster abundance at
zero redshift, ie., $\sigma_{8}=0.55\Omega_{\circ}^{-0.6}$ (Eke, Cole, \& Frenk 1996),
while the fifth is COBE normalized with $\sigma_{8}=1.35$.}
\tabcolsep 13pt
\begin{tabular}{ccccc} \hline 
Model & $\Omega_m$ & $\Omega_{\Lambda}$ & $\Gamma$ & $\sigma_{8}$ \\ \hline
$\Lambda_{\rm CDM1}$ & 0.3 & 0.7 & 0.25 & 1.13 \\ 
$\Lambda_{\rm CDM2}$ & 0.5 & 0.5 & 0.25 & 0.83 \\
$\tau_{\rm CDM}$ & 1.0 & 0.0 & 0.25 & 0.55 \\
O$_{\rm CDM}$ & 0.5 & 0.0 & 0.25 & 0.83 \\
C$_{\rm CDM}$ & 1.0 & 0.0 & 0.5 & 1.35  \\
S$_{\rm CDM}$ & 1.0 & 0.0 & 0.5 & 0.55 \\ \hline
\end{tabular}
\end{table}

For each cosmological model we average results over
10 nearly independent mock PSCz catalogues extending out to a 
radius of 170 $h^{-1}$Mpc which were produced by Enzo Branchini (see
Branchini et al 1999). Good
care was taken to center the catalogues to suitable LG-like observers
(having similar to the observed Local Group velocity, shear and
overdensity). Note that due to our ignorance in
assigning galaxy formation sites to the DM halo distribution and 
for a consistent treatment of all models, a biasing factor of 1 has 
been used in generating the PSCz look-alikes.

We analyse the mock PSCz catalogues density fields 
(with $R_{sm}=5$ $h^{-1}$ Mpc)
in the same fashion as that of the observed PSCz catalogue.
We find that the void shape distribution cannot discriminate among the
different models, a result which is similar with the shape-spectrum
analysis of superclusters (see Basilakos et al 2001). 

In figure 3 we plot the PSCz void-size frequency distribution
in logarithmic bins and compare it with the outcome of
all six models (hatched regions).
We see that the models with similar value of $\sigma_{8}$ produce
similar void-size distributions. We have verified this by using a
Kolmogorov-Smirnov test on the unbinned distribution of 
void-sizes from all 10 realizations. Indeed, all pairs of
model void-size distributions 
are excluded from being drawn from the same parent population 
at a high significance level ($>99.9\%$), except for
pairs that have similar values of $\sigma_{8}$, ie., 
$\Lambda_{\rm CDM1}$-C$_{\rm CDM}$, $\Lambda_{\rm CDM2}$-O$_{\rm CDM}$
and S$_{\rm CDM}$-$\tau_{\rm CDM}$, which appear absolutely consistent
among them. 

\begin{figure}
\mbox{\epsfxsize=9cm \epsffile{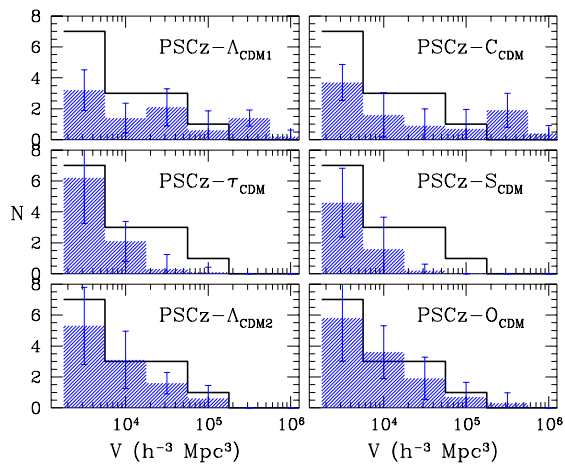}}
\caption{Void-size distribution comparison between the PSCz 
and look-alikes generated from 6 cosmological 
models. The model results are represented by the hatched regions and
the errorbars represent the scatter among the 10 realizations of
each model.}
\end{figure}

Since in the model-data comparison we deal with only one realization of
the Local Void distribution and since we have only 14 such voids, a
rather small number for the K-S test to give reliable results (as we
have verified using the K-S test with single model realizations), we
quantify the differences between models and data by performing a standard 
binned $\chi^{2}$ test, taking into account the scatter from the
10 different PSCz-like realizations. 
These probabilities are presented in table 4. Note the limitations of
this test due to the large covariance between bins. However, by
applying it to the model distributions, for which we have reliable
KS probabilities, we can assess its reliability.

Comparing the void-size distributions of the models 
among themselves we see that the $\chi^2$ test is less sensitive to
differences among the models with respect to the K-S test.
With the exception of the C$_{\rm CDM}$ and $\tau_{\rm CDM}$ 
models, which cannot be discriminated against, 
the other models follow the overall behaviour
found by the K-S test when it was applied to the voids 
of all 10 realizations.
\begin{table}
\caption[]{$\chi^2$ probability of significant 
differences between the model-model or
model-data void-size distributions. Each probability value corresponds to the
pair formed between the indicated model or data in the 
first column and the first row. In bold we indicate the
results of the PSCz-model comparison.}
\tabcolsep 3pt
\begin{tabular}{cccccccc}  \\ \hline
             & $\Lambda_{\rm CDM1}$  & C$_{\rm CDM}$  
& $\tau_{\rm CDM}$ & S$_{\rm CDM}$ & $\Lambda_{\rm CDM2}$ & O$_{\rm CDM}$ 
\\ \hline
{\bf PSCz}           & {\bf 0.002}& {\bf 0.006} & {\bf 0.001} & {\bf 0.000} & {\bf 0.195} & {\bf 0.870}\\
$\Lambda_{\rm CDM1}$ &           & 0.797 & 0.008 & 0.042 & 0.020 & 0.053\\
C$_{\rm CDM}$        &           &       & 0.114 & 0.428 & 0.221& 0.302\\
$\tau_{\rm CDM}$     &           &       &      & 0.906 & 0.166 &0.408\\ 
S$_{\rm CDM}$        &           &       &      &      & 0.007 &0.014\\
$\Lambda_{\rm CDM2}$ &           &       &      &      &    & 0.986 \\ \hline
\end{tabular}
\end{table}

Regarding the comparison between the PSCz and model voids, 
we see from both figure 3 and table 4 that 
the only models that fit the PSCz data, 
at a high significance level,
are models with $\sigma_{8}=0.83$; ie., the $\Lambda_{\rm CDM2}$ 
and O$_{\rm CDM}$, confirming the visual impression. 
Equivalent results are found by comparing
the largest or most extremely shaped void 
of each realization with that of the PSCz data.
This is not to say that an $\Omega_{\circ}=0.5$ should 
be preferred but rather that a $\sigma_{8} \simeq 0.83$ CDM model 
is consistent with the data (similar results are found from the whole
{\em pdf} study; Plionis \& Basilakos 2001). 

\section{Conclusions}
We have studied the properties of voids detected in the 
smoothed PSCz galaxy density field as connected regions under
some overdensity threshold.
We have investigated the biases that enter
in the detection of voids using our procedure and the PSCz 
smoothed density field. We reliably detect 14 voids within 80 
$h^{-1}$ Mpc, in the $R_{sm}=5$ $h^{-1}$ Mpc
smoothed field, having a median semi-major axis of $\sim 20$ $h^{-1}$ Mpc
and 8 voids in the relatively more distant Universe 
($r \le 130$ $h^{-1}$ Mpc) for the $R_{sm}=10$ $h^{-1}$ Mpc smoothed field,
having a median semi-major axis $\sim 28$ $h^{-1}$ Mpc.
Finally, we have compared our PSCz void-size distribution with the
corresponding ones generated from six cosmological
models and we find
that the CDM models that best reproduce the PSCz results
are those with $\sigma_{8} \simeq 0.83$, independent of cosmology.

\section* {Acknowledgments}
S.Basilakos has benefited from discussions with P.Coles. 
M.Plionis acknowledges the hospitality of the Astrophysics Group
of Imperial College. We both thank E.Branchini for providing us with
his reconstructed PSCz galaxy distribution and the PSCz look-alikes.
We both thank the referee, M.Vogeley, for helpful comments and suggestions.
\end{document}